\documentclass[a4paper,10pt]{article}

\usepackage[utf8]{inputenc}
\usepackage{fancyhdr}
\usepackage{authblk}
\usepackage[a4paper]{geometry}

\usepackage{hyperref}

\usepackage{cite}
\usepackage{amsmath,amssymb,amsfonts}
\usepackage{algorithmic}
\usepackage{graphicx}
\usepackage{textcomp}
\usepackage{xcolor}
\usepackage{flushend}
\usepackage{siunitx}
\fancypagestyle{firstpagefooter}{%
  \fancyhf{}
  
  \fancyfoot[L]{Accepted version of the paper submitted to IEEE SusTech, 2022\\
  DOI: \href{https://dx.doi.org/10.1109/SusTech53338.2022.9794179}{10.1109/SusTech53338.2022.9794179}\\978-1-6654-0664-2/22/\$31.00 \copyright~2022 IEEE}
}

\title{Enhancing the Performance of Multi-Agent Reinforcement Learning for Controlling HVAC Systems}
\author[1]{Daniel Bayer\footnote{Email address: daniel.bayer@fau.de}~}
\author[1]{Marco Pruckner\footnote{Email address: marco.pruckner@fau.de}}
\affil[1]{Energy Informatics, Computer Science 7, Friedrich-Alexander-University Erlangen-N\"urnberg}
\date{}

\begin{document}

%\IEEEauthorblockA{\textit{Energy Informatics} \\
%\textit{Computer Science 7} \\
%\textit{Friedrich-Alexander-University Erlangen-N\"urnberg}\\
%Erlangen, Germany \\
%\{daniel.bayer, marco.pruckner\}@fau.de}}

\maketitle

\begin{abstract}
Systems for heating, ventilation and air-conditioning (HVAC) of buildings are traditionally controlled by a rule-based approach.
In order to reduce the energy consumption and the environmental impact of HVAC systems more advanced control methods such as reinforcement learning are promising.
Reinforcement learning (RL) strategies offer a good alternative, as user feedback can be integrated more easily and presence can also be incorporated.
Moreover, multi-agent RL approaches scale well and can be generalized. 
%For reducing the energy consumption of HVAC systems self-learning control algorithms based on reinforcement learning (RL) can be used, that get information collected by a smart building infrastructure.
%Generally, multi-agent RL approaches promise sustainability as they scale well and can be generalized.
%An popular instance of RL is Deep-Q-Learning, but it can be slow and hard to train in a multi-agent setting, especially for controlling HVAC systems.
In this paper, we propose a multi-agent RL framework based on existing work that learns reducing on one hand energy consumption by optimizing HVAC control and on the other hand user feedback by occupants about uncomfortable room temperatures.
Second, we show how to reduce training time required for proper RL-agent-training by using parameter sharing between the multiple agents and apply different pretraining techniques.
Results show that our framework is capable of reducing the energy by around 6\% when controlling a complete building or 8\% for a single room zone.
The occupants complaints are acceptable or even better compared to a rule-based baseline.
Additionally, our performance analysis show that the training time can be drastically reduced by using parameter sharing.
\end{abstract}

\thispagestyle{firstpagefooter}

%\begin{IEEEkeywords}
%Multi-Agent Reinforcement Learning, Independent Q-Learning, Shared Parameters, HVAC Systems, Building Controls, Energy Efficiency, Energy Saving Controls
%\end{IEEEkeywords}

\section{Introduction}
\label{sec:introduction}
Several studies show that buildings require a huge amount of energy for heating and cooling and thus cause a significant proportion of global CO$_2$ emissions.
According to the Building Global Status report \cite{2018_Building_GlobalStatusReport} buildings are accountable for about 39\% of the global CO$_2$ emissions in the year 2017.
The government of Germany estimates that heating, cooling and lightning of buildings including indirect emissions are responsible for 30\% of the CO$_2$ emissions of the whole country \cite{2020_KlimaschutzInZahlen_EN}.
It is estimated that the U.S. building sector accounts for around 29\% of total U.S. CO$_2$ emissions when including indirect emissions as well, with office buildings emitting half of this \cite{2018_Leung_DecarbonizingUSBuildings}.

Thermal comfort and indoor air-quality for all kinds of buildings are provided by so-called heating, ventilation and air-conditioning (HVAC) systems.
HVAC systems are mainly responsible for the energy consumption and CO$_2$ emissions of buildings.
Up to now, most of these systems are controlled by a static, rule-based control \cite{2021PredictiveControlForBuildings} meaning mainly simple rules which are based on some predefined measured data and heuristics \cite{2011AdvancedHVACControl}.
%As the name suggest, an engineer (preferably with a long experience in this sector) defines a rule that is based on some predefined measuring data and may utilize heuristics .

In order to reduce the CO$_2$ emissions by saving energy and enhancing sustainability of buildings better methods for the control of these systems are needed.
Further optimization with more advanced methods other than rule-based control like model-predictive control (MPC) or reinforcement learning (RL) seems to be promising in the context of saving energy in the building sector.
RL in combination with deep learning (DL) can outperform hand-crafted controls as it can detect complex structures and learn automatically.
Both methods, MPC and RL have been applied to control HVAC systems in recent years. Many publications show that these strategies can principally reduce energy consumption without reducing the comfort of the occupants \cite{2020BrandiDRLIndoorTempOpti,2021_Yu_ReviewDRLForSmartBuildingEnergyManagement,2021_Li_MultizoneThermalControl_MultiAgent_DRL}.
Generally, saving energy consumed by HVAC devices can be obtained in two ways: physically by e.g. improving building insulation, or technically by optimizing the control of the HVAC system.
The latter is the aspect we will focus on in this paper, i.e. we develop an energy saving control based on RL.
To break down the big context of HVAC systems to a feasible size, we address only heating of buildings in this work.
As heating air and water in buildings requires 71\% of consumed energy in the complete building sector in Germany \cite{2020_KlimaschutzInZahlen_EN}, heating seems to be most attractive for reducing energy consumption.
Smart heating is a term that is commonly used, especially in the context of home energy management systems (HEMS) which are used in residential buildings.
A good overview of HEMS and smart heating is given in Nacer et al. \cite{2017_Nacer_SmartHome_SmartHEMS_SmartHeating}.
%Our framework does not concentrate on residential buildings, we rather focus on office buildings. 

While MPC needs a mathematical model of the HVAC system, RL only learns from observations and thus is totally model-free.
Even though using RL seems to be very sustainable as it is easy and fast to apply, the process of learning from observations can be slow and thus creating a control based on RL might require much training time.

%As this does not meet the occupants wishes, it is the of finding a balance between these wishes and energy consumption. 
Therefore we propose a RL framework, that addresses sustainability in two ways.
On the one hand we enhance the sustainability of HVAC systems by optimizing their control.
On the other hand we enhance the sustainability of creating HVAC system controls.
To address both aspects, our RL framework is
\begin{itemize}
    \item scalable in terms of the number of controlled elements,
    \item generalizable, i.e. capable of using prior knowledge, and
    \item adaptable to individual zone conditions.
\end{itemize}

This paper is structured as follows:
%We finish this section with a short introduction of the fundamentals of HVAC systems and reinforcement learning.
In Section \ref{sec:related_work} we give a short review on related work and identify gaps in knowledge.
In the following Section \ref{sec:fundamentals_RL} we present the fundamentals of the proposed framework.
In Section \ref{sec:proposed_framework} we describe our RL-framework from a more theoretical point of view.
We proceed with an evaluation of this framework in Section \ref{sec:evaluation} and thereafter present results of energy reduction and thermal comfort in Section~\ref{sec:restults}.
Finally we summarize our paper and give a short outlook on further research questions in Section~\ref{sec:conclusion}.

\section{Related Work}
\label{sec:related_work}
There has been done a lot of research in the last three decades on reducing energy consumption of HVAC systems by optimizing its control.
% TODO: Frage: 2018MPCforEnhancingHVACEfficiency ist ein Literatur-Review, passt das so in diesem Kontext?
Approaches discussed in \cite{2018MPCforEnhancingHVACEfficiency,2014HVAC_with_MPC} try to outperform traditional rule-based control strategies by using MPC.
The idea behind controlling a process with MPC is to find an optimal control command for a system at every individual time step.
The optimal command at the current time step is found by predicting the future behavior for a given time horizon.
The future behavior of the controlled system is predicted using a given model of the system \cite{2018MPCforEnhancingHVACEfficiency}.
The authors of \cite{2014HVAC_with_MPC} are able to reduce the energy consumption by 9.4\% in the heating period for an office building.
The authors of \cite{2018MPCforEnhancingHVACEfficiency} estimate the savings potential even up to 20\%.
A major drawback of MPC is the need for a good mathematical model of the HVAC system, thus MPC is not generalizable to HVAC systems of other buildings.

In the last five years, controlling algorithms based on RL in combination with DL gained more and more attention in the context of controlling HVAC.
A recently published review paper by Yu et al. discusses 15 different references dealing with this topic \cite{2021_Yu_ReviewDRLForSmartBuildingEnergyManagement}.

% using so-called Deep-Q-Networks (DQN) is used 
\subsection{State of the Art}
In \cite{2020BrandiDRLIndoorTempOpti} Deep Q-Learning is used to control the heating for a single thermal zone.
The variable setpoint in this setting is the supply water temperature.
Training and evaluation takes place between November and December by using a self-modeled building.
As simulation environment the authors use EnergyPlus.
The reward function is a weighted sum of the supplied heating energy and the temperature control performance.
This means that temperature wishes of the occupants are not modeled directly, moreover a static band is defined in which the room temperature has to be if people are present.
As a result, the trained agent can save energy between 5\% and 12\% compared to the baseline and additionally reduce the sum of temperature violations by up to 50\%.

Q-Learning is also successfully applied during the summer months for 
controlling air conditioner in residential buildings by Lork et al. \cite{2020LorkRLFrameworkUncertAw}.
Besides this control, they also investigate uncertainties like changing occupancy, as they do not have access to smart building information like the current occupancy for a room.
In the end, it is possible to reduce the energy consumption by 3.6\% compared to a rule-based baseline while improving thermal comfort.

While \cite{2020BrandiDRLIndoorTempOpti} and \cite{2020LorkRLFrameworkUncertAw} use a single agent, the authors of \cite{2020MARCO} propose a multi-agent approach for controlling HVAC systems.
Every controllable element like central chillers or air-handling units are controlled by a single agent.
Even though, all agents get the same reward that is a sum of total energy consumption and thermal comfort.
Different to both previous works, this publication only focuses on cooling these thermal zones.
The goal of training is that the agents are able to govern the indoor temperature to stay in a predefined band.
This band depends on the room temperature and other factors like humidity or the mean radiant temperature.
%As in some other publications, EnergyPlus is used as simulation environment.
To speed up the training and ensure scalability the authors use transfer learning meaning that all agents are pretrained.
%This means that the agents are pretrained.
For this procedure all components have a rule-based control, except one, that is controlled by a learnable agent.
Thus, this agent can learn more easy. After the pretraining the other agents are initialized using a copy of the single agent.
Finally, the proposed framework can reduce the energy consumption by $17\%$ without reducing the thermal comfort compared to a self-defined rule-based baseline.

Another multi-agent approach is presented in \cite{2021MultiAgent_DRL_HVAC}, for which an actor-critic algorithm is used.
In contrast to the works mentioned before, every agent gets an individual reward but the reward is still a sum of energy consumption and thermal comfort.
As a result, these agents are able to keep temperature inside self-defined bands most of the time.
But the energy saving potential and thermal comfort for the occupants has not been evaluated directly.

A recent approach \cite{2021_Li_MultizoneThermalControl_MultiAgent_DRL} uses a RL-algorithm called multi-agent deep deterministic policy gradient (MADDPG) to control the temperature and humidity setpoint for each thermal zone in a given building.
Every agent controls an individual thermal zone, and receives an individually computed reward in contrast to \cite{2020MARCO}.
In the end, results show that about $15\%$ of energy are saved compared to a self-defined, rule-based baseline without reducing the thermal comfort.

%First literature reviews that also cover the topic of optimizing the control of HVAC were published recently \cite{2021_Yu_ReviewDRLForSmartBuildingEnergyManagement}.
%On the other hand, it turns out that publications with multi-agent approaches appear very rarely.

Up to now, the following gaps in knowledge can be identified:
There has been few research using multi-agent approaches in the context of HVAC using a shared reward, i.e. independent Q-learning. Especially it is not clear, how to properly train multiple agents to get a fast convergence of the rewards during training without having an unacceptable long training time.
Shared parameters have not yet been implemented in the context of HVAC control.
The target temperature for the computation of the thermal comfort is always modeled as a static band or inferred from created neural networks if people are present in the building \cite{2020MARCO,2020BrandiDRLIndoorTempOpti}. The occupants never have the ability to express their comfort temperature wishes.
There are only few frameworks for simulating a complete HVAC system with occupancy as well.
%Most of the approaches \cite{2020BrandiDRLIndoorTempOpti,2020LorkRLFrameworkUncertAw,2020RLWholeHVAC} can only be used for the building they have been trained on, thus transferring the model to a new building requires new training from scratch.
%There is a lack of generalizeable models.
% TODO: Optional:
% Generalizability is important, as transfer leraning in this context has already been applied succesfully

\subsection{Our contribution}
Having the knowledge gaps in mind, this work presents a framework for energy reduction in office buildings that fulfills the following goals:
\begin{itemize}
    \item \textbf{Scalability and sustainability:} 
	Adding new thermal zones or other controlled components has to be simple and computational efficient.
	A retraining of the complete model has to be prevented to save run time during training.
	\item \textbf{Generalizability:}
	The individual models should generalize to some extend. This means that transferring existing models to new buildings or newly added components should be possible.
	\item \textbf{Adaptability to individual room conditions: }
	Almost every building has rooms that are heated up faster by the sun than others.
	Besides of the sun, also occupants might have different wishes regarding the room temperature.
	It is easy to think of situations where in one room people wish a lower temperature than in another.
	%NEW NEW NEW
	In our situation the occupants can send a complaint to the system if their comfort temperature is not within a band of $\pm$\SI{1.0}{\degreeCelsius}.
	%END END END
	The framework should be able to pay attention to these individual room conditions.
\end{itemize}
Moreover, we evaluate the impact of various hyper-parameters like epsilon decay, the time resolution or reward scales.
We use a multiple agent variant of Q-Learning, called independent Q-Learning and approximate the Q-function by Deep-Q-Networks.

The framework proposed in \cite{2020MARCO} is closest to our contribution.
Compared to that, our work heavily differs in the way we simulate the occupancy, the way we compute the reward for the RL-algorithm and we pay more attention to the theoretical background of mutli-agent RL using a shared reward.
Moreover our simulation concentrates on the heating period, while the use case discussed in \cite{2020MARCO} considers summer time for cooling.
%Compared to \cite{2020BrandiDRLIndoorTempOpti} we use a multi-agent approach and control a complete building.
% we allow inidividual temperature wishes

%
% Theoretisch koennten hier auch die fundamentals stehen
%

\section{Fundamentals of Reinforcement Learning}
\label{sec:fundamentals_RL}
In this section we give a short introduction to RL based on Sutton et al. \cite{SuttonBartoRLIntro}.
RL consists of four components: An environment, one or more agents that can manipulate the environment,  the current state of the environment for every time step $t$ encoded as vector $s_t$ as well as a reward signal $r_t$.
The reward is a real number indicating the benefit or badness of the current state.
The action an agent takes at time step $t$ is denoted as $a_t$.

The goal of the learning process for the agents is to find a strategy that maximizes the cumulative, discounted reward, called overall return:
\begin{equation}
    \label{eq:overall_return}
    G_t := \sum_{k = 0}^{\infty} \gamma^k R_{t+k} \Longrightarrow G_t = R_{t} + \gamma G_{t+1}
\end{equation}
where $R_i$ denotes the random variable of the reward at time step $i$ and $\gamma \in (0,1)$ is a predefined discount factor.
The agent's strategy is called policy.

\subsection{Markov Decision Processes}
A classical formalization of making decisions sequentially is to model it as a Markov Decision Process (MDP).
As the name Markov suggests, we assume the current state only depends on the last state, but not on those that were further back.
% write something about the global reward, v^* and q^*
May $S_t$ and $A_t$ be the random variables of the state and the action at step $t$.
For the further course we introduce the value function $v_\pi(s)$ as the expectation of the overall return by following policy $\pi$ starting in state $s$, in terms:
\begin{equation}
    v_\pi(s) := \mathbb{E}_\pi\left[G_t\,\mid\,S_t = s\right]
\end{equation}
and the action-value function $q_\pi(s)$ as the expectation of the overall return by following policy $\pi$ starting in state $s$ and choosing action $a$ there, in terms:
\begin{equation}
    \label{eq:q_function}
    q_\pi(s,a) := \mathbb{E}_\pi\left[G_t\,\mid\,S_t=s,\; A_t = a\right]
\end{equation}
Bellman noticed that a policy $\pi^*$ that is optimal in terms of the value-function is also optimal for the action-value function \cite{1957BellmanMDP}.
Given such an optimal policy $\pi^*$ one can write using \eqref{eq:overall_return}
\begin{equation}
    \label{eq:bellman_opti_q}
    q_{\pi^*}(s,a) = \mathbb{E}\left[R_{t+1} + \gamma v_{\pi^*}(S_{t+1})\,\mid\,S_t=s,\,A_t=a\right]
\end{equation}
or rewrite it to
\begin{equation}
	q_{\pi^*}(s,a) = \mathbb{E}\left[R_{t+1} + \gamma \max_{a'} q_{\pi^*}(S_{t+1}, a') \,\mid\,s,\,a\right]
	\label{eq:2_bellman_opti_q}
\end{equation}
where $a'$ denotes the next action.

\subsection{Q-Learning}
Q-Learning is a well-known algorithm in the context of RL that directly utilizes \eqref{eq:2_bellman_opti_q}, first introduced in the late 1980s in \cite{1989QLearning}.
It has been successfully applied in many different environments like playing games.
In recent years it has also been used successfully in the field of controlling HVAC systems \cite{2020BrandiDRLIndoorTempOpti, 2020LorkRLFrameworkUncertAw}.
% TODO: present basic update rule
The basic idea behind Q-Learning is to obtain a good approximation of \eqref{eq:2_bellman_opti_q}, by an iterative application of the update rule
\begin{equation}
    Q(s,a) \leftarrow Q(s,a) + \alpha \left( r_t + \gamma \max_{a'} Q(s',a') - Q(s,a) \right)
    \label{eq:Q_learning_update_rule}
\end{equation}
where $r_t$ is the current reward.
For a given state $s$ the agent takes the action $a$ that maximizes $Q(s,a)$.
During the training period we also want to explore apparently bad actions, thus the agent takes a random action with a probability of $\epsilon$. Thus Q-learning is also referred as off-policy algorithm.

\subsection{Deep-Q-Learning using Deep-Q-Networks}
The Q-function \eqref{eq:Q_learning_update_rule} can have various representations, for example it could be a traditional table.
Nowadays it is approximated using a neural network, which is called Deep-Q-Network (DQN).
The algorithm for training such a DQN was first proposed in \cite{2015DQNBasics} and is called Deep-Q-Learning.
Because classical Deep-Q-Networks tends to overestimate the Q-values, \cite{2016DDQN} introduces a Double Deep-Q-Network DDQN.
These networks replace the direct maximization in \eqref{eq:Q_learning_update_rule} by consulting the Q-network again.

Extending DQN or DDQN for a multi-agent setting can be done straight forward by approximating multiple Q-functions by having multiple DQNs or DDQNs in parallel \cite{2015_MultiAgent_Cooperation_Competition_DRL}.

%
% Frage: Steht irgendwo, dass DQN bei anderen Papern bereits gut performen?
% Und warum verwenden wir DQN und nicht DDPG? (wir sehen hierin keinen Mehrwert)
%

\section{Methodology and Proposed Framework}
\label{sec:proposed_framework}
%We want to optimize the heating control of a office building by using reinforcement learning.
%Therefore, we approximate the Q-Function by a neural network using DQNs.
In the following, we explain our framework, which is used to optimize the heating control of an office building and define the state space, action space and reward function.
Additionally we describe the pretraining, parameter sharing and the implementation.
We will use the words zone, thermal zone and room as synonyms because we use a building in the further evaluation that fulfills this situation.
Our framework uses a time resolution of 1 hour.
As one episode we define one month.

\subsection{State Space}
The state has to include all environmental parameters that have a notable impact on the climatic conditions inside a building.
We select the outdoor air temperature besides direct and indirect solar radiation as these variables have a deep impact on the indoor conditions.

Time information is important for planning, e.g. for the question when to increase zone temperature setpoints in winter, even if no people are present at the moment.
While some publications do not include time information in their state like \cite{2020LorkRLFrameworkUncertAw}, other papers claim that to be a classical approach in literature \cite{2020BrandiDRLIndoorTempOpti}.
In this work the state contains the current time of day, the day of the week and the calendar week.

Finally, we extend the state by adding important information about every thermal zone.
So we use the current zone temperature, the current relative occupancy and the prediction of the occupancy over a time horizon of two hours.
The relative occupancy is computed by the current occupancy divided by the maximal amount of people that can be placed in the zone.

The state is encoded as a vector of size $6 + n_{zones} \cdot 4$ where $n_{zones}$ is the number of existing zones.
An overview of the information encoded in the state can be found in Table \ref{tab:state_components}.

\begin{table}[htbp]
\caption{Elements of the state vector}
\begin{center}
\begin{tabular}{|ll|r|r|}
\hline
\multicolumn{2}{|l|}{\textbf{Variable description}} & \textbf{Min.} & \textbf{Max.}\\
\hline
\multicolumn{4}{|l|}{\textbf{\textit{General variables}}}\\
\hline
& Day of Week             & 0.0 & 6.0     \\
& Minutes of Day          &  0.0& 1439.0  \\
& Calendar Week           &  1.0&   53.0  \\
& Outdoor Air Temperature  [\textdegree C] &-20.0&   40.0 \\ 
& Direct Solar Radiation   [Wm$^{-2}$] &  0.0& 350.0  \\
& Indirect Solar Radiation [Wm$^{-2}$] &  0.0&1000.0  \\
\hline
\multicolumn{4}{|l|}{\textbf{\textit{Variables for every zone}}}\\
\hline
& Zone Temperature [\textdegree C] & 10.0 & 34.0  \\
& Zone Relative People Count &0.0 &   1.0 \\
& Zone Rel. People Count in 1 Hour &0.0 &   1.0 \\
& Zone Rel. People Count in 2 Hours &0.0 &   1.0\\
\hline
\end{tabular}
\label{tab:state_components}
\end{center}
\end{table}

By giving this definition of the state space we can ensure that a state vector at time step $t$ encodes all important information.
Thus the agents do not have to record the last state vectors and the Markovian assumption holds, which is essential for Q-learning.

\subsection{Agent design and action space}
To address our goal of scalability we decide to use a multi-agent approach as adding new components should be easy.
If there were one big central agent that holds one big DDQN, one could not add more controlled variables or add more actions as this would require adding more input or output layers to the DDQN.

In our setting one agent controls one thermal zone, where the heating setpoint has to be controlled.
The heating setpoint in each zone can be adjusted in a range from \SI{15.0}{\degreeCelsius} to \SI{25.0}{\degreeCelsius} in \SI{1.0}{\degreeCelsius} steps.
So in every step one agent can choose one out of 11 actions. 

As the state vector contains the information of all thermal zones, a removal of the variables of the uncontrolled zones is required in every step.
As a result every agent takes a vector of size 10 as input.

Please note that all agents control the same kind of variable, i.e. heating setpoints.
Moreover, they all get the same type of input parameters, just with different values for the different zones.
% TODO: Hier könnte man nun schreiben:
% Thus it is natural to develop some pretraining techniques

The DDQNs are implemented as fully connected neural networks with two hidden layers.
Between the hidden layers we introduce special layers for layer normalization.
This normalization technique reduces the training time \cite{2016LayerNormalization}.

\subsection{Reward}
The reward function is driven by the design goal that the agents should adapt to the occupants comfort temperature, that is expressed by complaint messages if the comfort temperature is not met.
If there are more occupants in one zone, we average the occupants mean temperatures.
This seems sensible for office buildings, because occupants in an office will likely discuss about their comfort temperatures.
One can expect that they will agree on the average comfort temperature of all occupants.
To address the complaints for the RL-algorithm, we have to define a measure of these complaints.
May $Z$ be the set of thermal zones with $|Z| = n_{zones}$, $t$ the current time step, $T_t(z)$ the current temperature in zone $z$ at time $t$ and $S_t(z)$ the mean comfort air temperature of the occupants in zone $z$ at time $t$.
The difference between the desired and the actual temperature for zone $z\in Z$ at time $t$ is expressed as
\begin{equation}
    \label{eq:reward_fn_part_mstpc}
    \delta_t^{Temp}(z) :=
    \begin{cases}
		0 & \text{if } z \text{ is unoccupied} \\
		| T_t(z) - S_t(z) |& \text{otherwise}
	\end{cases}
\end{equation}
Then, we can define the manual setpoint complaint magnitudes (MSTPC) per zone $z$ at time $t$ as
\begin{equation}
    \label{eq:mstpc_per_zone}
    \forall z \in Z: \bigtriangleup_t(z) =
	\begin{cases}
		0 & \text{if } \delta_T(z) \leq 1\\
		\delta_t^{Temp}(z) & \text{otherwise}
	\end{cases}
\end{equation}

Besides fulfilling the occupants wishes, saving energy is the second, probably contradicting goal.
May $E_t(d)$ be the energy consumption in kWh from time step $t-1$ to the current time step $t$ for a device $d$.
The energy consumption of the complete building's HVAC system is thus the sum of $E_t(d)$ over all devices $d$ of the HVAC system $D$.
So we finally define our reward at step $t$ as the negative value of a weighted sum of both aspects%\vspace{1mm}
\begin{equation}
    \label{eq:reward}
    r_t^E := -\left(\;\lambda_{e} \sum_{d\in D} E_t(d) + \lambda_m \sum_{z\in Z}\bigtriangleup_t(z)\;\right)
\end{equation}
where $\lambda_{e}$ and $\lambda_m$ are constant factors for scaling.
Because sudden peaks in the energy consumption might overweight the thermal comfort aspect, we slightly modify this reward.
Therefore we flatten energy consumption values $E_t^{all} := \sum_{d\in D} E_t(d)$ that are above the a given level $l_{clip}$:
\begin{equation}
    \label{eq:e_flat}
	E^{Flat}_t := \begin{cases}
			E^{all}_t & \text{if } E^{all}_t \leq l_{clip}\\
			(E^{all}_t-l_{clip})\cdot 0.1 + l_{clip} & \text{otherwise}
	\end{cases}
\end{equation}

Even though every agent has its own Q-Network, this reward is defined on a building level.
In literature, such a situation is referred as independent Q-learning \cite{1993_MultiAgentRL_IndepVsCoopAgents}.
This is the most popular approach to multi-agent RL and has proven to work fine in many cases \cite{2017_MulitAgent_RL_with_ExperienceReplay}.
In our setting it is not possible to calculate an exact reward on the level of agents, as a typical HVAC system has many components (like a central heating unit) whose energy consumption strongly depends on the decisions that are made somewhere else (like the actual room temperature).
Modeling these coherences would be very time consuming and would contradict to our goal of creating a generalizable and sustainable framework.

Our definition of the reward \eqref{eq:reward} is very similar to that of \cite{2020MARCO} and \cite{2020BrandiDRLIndoorTempOpti}.
Our defintions stands out from the others in directly integrating the occupants' feedback.
Compared to \cite{2021_Li_MultizoneThermalControl_MultiAgent_DRL} we define the reward on building level.

\subsection{Pretraining}
The expected run time for training a multi-agent approach is higher than for an individual agent due to two reasons. First, the action space, that has to be searched, becomes larger.
Second, there are more agents for which inference and backpropagation has to be calculated.
Thus, we develop two pretraining techniques that counteract longer run time.

The first pretraining technique, a single agent in broadcasting mode mainly addresses the first reason for longer training.
By default, we have $n_{zones}$ agents that control $n_{zones}$ thermal zones.
In this case we just use the first agent, and delete the other $n_{zones} - 1$ agents.
The control commands of the first agent are now sent to all thermal zones, i.e. the commands of this agent are broadcasted to all thermal zones.
Thus, we can dramatically reduce the number of possible actions combinations.
After the pretraining with one single agent in broadcasting mode, we initialize the other agents as well. The new agents get a copy of the network from the first agent.

The second pretraining technique, a partially rule-based control addresses both reasons for longer training and is based on \cite{2020MARCO}.
This case is almost like the previous one, but instead of broadcasting the commands we use a rule-based control for the other agents.
This means that we just use the first RL-agent, but now replace the other $n_{zones} - 1$ RL-agents by rule-based agents.
Thus, we save a lot of time for backpropagation and have a small action space too.
After the pretraining with one RL-agent, we initialize the other agents as well. As in the previous case, the other agents get a copy of the network of the first agent.

\begin{figure}[htbp]
	\centerline{\includegraphics[scale=1, width=0.55\textwidth]{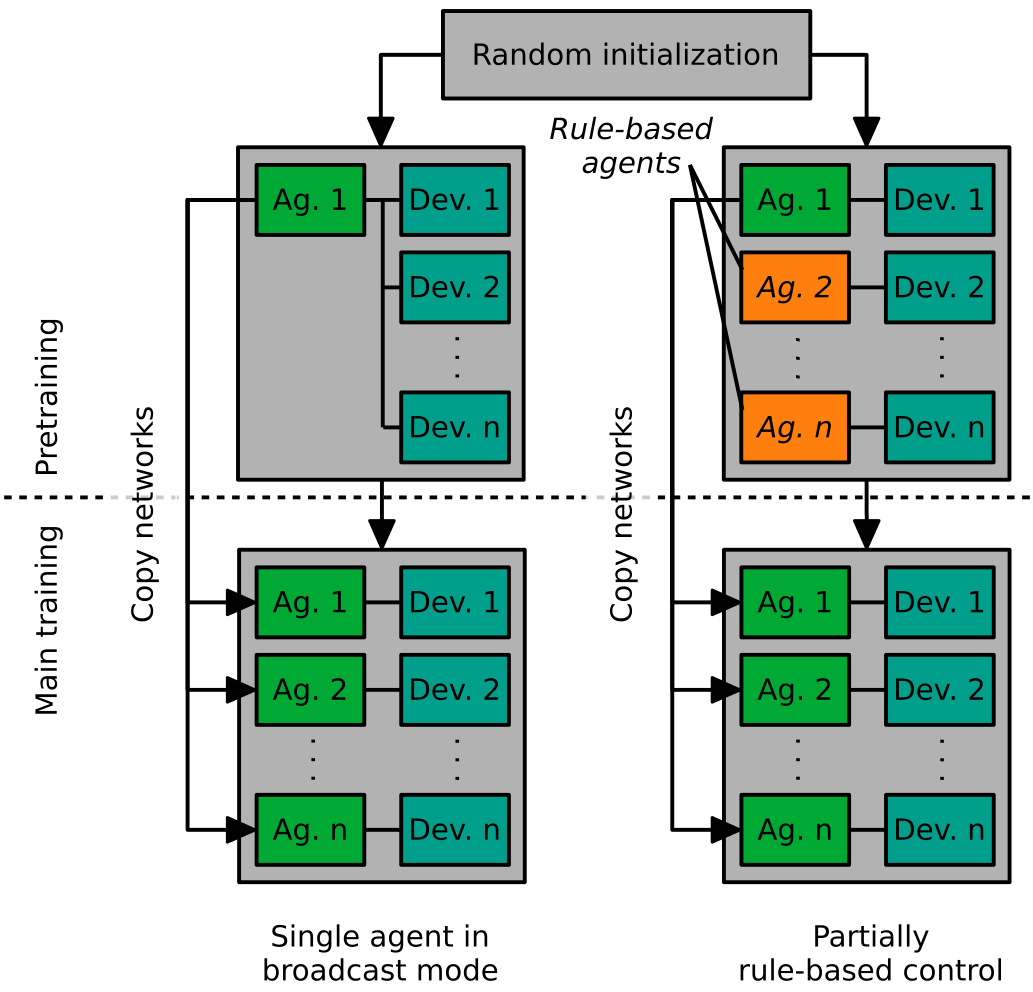}}
    \caption[Agent pretraining overview]{Schematic overview of the two pretraining techniques.}
    \label{fig:pretraining_overview}
\end{figure}

Fig. \ref{fig:pretraining_overview} shows a schematic overview of these two pretraining techniques.

\subsection{Semi-multi-agent learning using parameter sharing}
After pretraining has been finished, it still remains important to reduce the training time.
In our setting all agents should roughly learn the same policy.
These policies should only differ slightly because of different zone settings, like window fronts etc.
Therefore, it makes sense to share the networks of the agents during the main training.
Just for the last few episodes we allow the agents to have individual networks.
% TODO: hier koennte man noch darauf eingehen, dass:
% We can easily generalize the networks (as requested by the goals of the framework, see \cref{sec:goals_of_implementation}), because there is just one per device class.

In literature it has often been proven that parameter sharing reduces training time and enhances the training results, as the agents can share their knowledge \cite{2021_Gronauer_MultiAgentDRL_Review}.

In fact one can see this as a mixture of a single and a multiple agent setting.
A multiple agent setting, because the agents can still make individual decisions due to the fact that they obtain different subsets of the current state.
On the other hand we can see this as a single agent setting, because all information flows together into one Q-network.

\subsection{Simulation and framework components}
Our framework is implemented in Python and consists of several classes, which we will briefly introduce.
The central controller class holds the DQN/DDQN-algorithm and is enhanced by additional commands that are required for controlling the simulation.
The agent class holds the implementation of the neural networks, that are based on PyTorch.
Moreover, there is a rule-based implementation which is required for the baseline and a pretraining technique.
The building class initializes and manages the connection to the simulation software EnergyPlus \cite{2000EnergyPlusASHRAEJournal}.
As an interface between EnergyPlus and our code we use COBS \cite{2020COBSPosterAbstract}.
The occupancy controller class holds a schedule for a week defining the occupancy for every zone at every time step including the temperature wishes of the occupants.
The schedule can be randomly created or loaded from a file for making results comparable.
Finally, there is a training controller class for initialization of the complete framework, a replay buffer class required for the DDQN/DQN-algorithm and a class for piping the output to an SQL database.

The classes of the building, central controller, occupancy controller, replay buffer, SQL-ouput and the training controller are singletons.
The implementation is also available for the public on GitHub\footnote{https://github.com/danielrenebayer/HVAC-RL-Framework}.

\subsection{Baseline}
We define a rule-based control strategy as a baseline for comparison with the RL-training and for pretraining the partially rule-based control.
For the baseline we use one agent per controlled zone, which is able to control the zone heating setpoint.
Anyway, all the other components (e.g. central boiler) are controlled by a rule-base strategy.
For these components we do not define an separate agent.

The building is heated for the baseline from Monday to Friday between 7:00~am and 6:00~pm. During this time the heating setpoint is set to \SI{21.0}{\degreeCelsius}, outside this period it is set to \SI{15.0}{\degreeCelsius}.
The choice of \SI{21.0}{\degreeCelsius} fits the comfort temperature averaged over all occupants.

\section{Evaluation}
\label{sec:evaluation}

\subsection{Simulation setting}
\begin{figure}[htbp]
	%\centering
	%\includegraphics[]{figures/building_3D}
	\centerline{\includegraphics[width=0.55\textwidth]{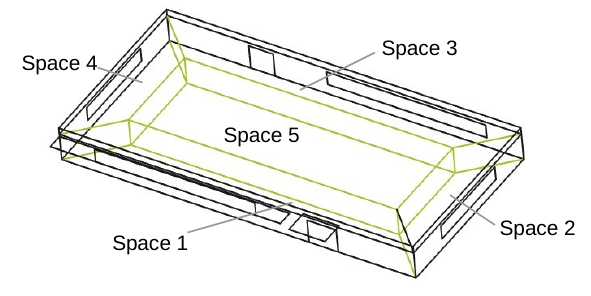}}
	\caption[3D plot of the building]{3D plot of the building. Green lines represent interior walls. On the south side there is a overhang on the building to shield the sun. The visualization is exported from EnergyPlus, similar to \cite{2020RLWholeHVAC}.}
	\label{fig:building_3D}
\end{figure}

The used building for analysis comes along with EnergyPlus as one of many sample buildings.
The building has five rooms named \textit{Space 1} to \textit{Space 5}, each of them is an individual thermal zone. Space 1 to 4 are the elongated rooms with windows, Space 5 is the central interior. 
The height of the building is \SI{3.05}{\metre}. The width is \SI{30.5}{\metre}, the length is \SI{15.2}{\metre}.
Fig. \ref{fig:building_3D} shows a 3D-visualization of the building.
The same building is also used in literature, for example in \cite{2020RLWholeHVAC}.
% FRAGE: wir haben diese Quelle bisher nicht erwähnt, ist dies ein Problem?

%Commonly, office buildings have a centrally controlled heating, ventilation and air-conditioning (HVAC) system.
%Thus, implementation of energy saving strategies is more simple compared to residential buildings.

The training takes place with weather data for Fairbanks for the complete month of January.
The values of the hyper-parameters used for training the DDQNs can be found in Table~\ref{tab:hyperparameters}.
We start with an exploration rate $\epsilon$ of 100\% and decay it linearly down to a final value of 5\%.
As optimizer we use the popular Adam \cite{2015Adam}.
The learning rate is decayed stepwise from 0.8 to 0.01 at the end.
The replay buffer size is set to store the last 576 values, as this corresponds to a history of 24 days.
The training has generally three phases:
First, we do an optional pretraining for \num{12500} episodes using one of the above defined techniques.
This is followed by the main training, which takes \num{12500} episodes (or \num{25000} episodes without pretraining) again.
During the main training phase we use parameter sharing.
Finally we do an individual training without parameter sharing for 250 to 1000 episodes.

For modelling the occupancy we define the center room Space 5 and one side room (Space 3) to be the main office rooms with presence between 8:00 am and 4:00 pm.
The other rooms are used as conference rooms with changing occupancy over the week.
Still, the schedule repeats for every week, except holidays without presence of anyone.
The occupants of the conference rooms have different temperature wishes during the week.

% TODO:
% fill in values!
\begin{table}[htbp]
\caption{Overview of the hyper-parameters used for training}
\begin{center}
\begin{tabular}{|l|c||l|c|}
\hline
\textbf{Parameter} & \textbf{Value} & \textbf{Parameter} & \textbf{Value} \\
\hline
discount factor $\gamma$ &  0.9& train. episodes     & 12500 \\
replay buffer size       & 576 & $\lambda_e$           & 0.008 - 0.004 \\
batch size               & 256 & $\lambda_m$           & 0.06 - 0.12 \\
target network upd. freq.& 2   & time resolution & 1 h \\
learning rate $\alpha_{start}$ & 0.8  & $\epsilon_{start}$ & 1.00 \\
learning rate $\alpha_{final}$ & 0.01 & $\epsilon_{final}$ & 0.05 \\ 
\hline
\end{tabular}
\label{tab:hyperparameters}
\end{center}
\end{table}

\subsection{Training time per episode}
%A longer training time for multi-agent DQN can be expected as the optimal Q-function, that is directly received from \eqref{eq:2_bellman_opti_q}, changes:
%\begin{equation}\begin{split}
%\label{eq:3_independent_q_learning}
%q_*(s,a) &= \mathbb{E}\left[ R_{t+1} + \gamma \cdot \max_{a'} q(S_{t+1}, a') \right]= \\
%&=\sum_{s',r,\overline{a}} p(s',r | s,a,\overline{a}) \left[ %r+\gamma\cdot\max_{a'} q_*(s',a')\right]
%\end{split}\end{equation}
%where $\overline{a}$ are the actions of all other agents. Thus, we have to wait longer for the Q-values to converge.
A longer training time for multi-agent DDQN/DQN can be expected for two reasons.
First, from a global perspective, there are more possible actions aggregated over all agents.
Second, one episode of training takes longer as there are more agents for inference and learning.

We can see the latter aspect in \cite{2020RLWholeHVAC}, where the run time for on episode (roughly $36$ seconds) is the limiting factor for the number of training episodes.
Our framework is much faster, especially if we use parameter sharing.
In the last case our mean runtime per episode is about $4.8$ to $5.0$ seconds.
Without shared networks the mean runtime per episode is about $10.0$ seconds.
%\begin{figure}[htbp]
%	\centerline{\includegraphics[scale=1, width=0.5\textwidth]{figures/runtime_overview}}
%    \caption[Run time overview]{Overview of the runtime in different settings compared to a value from literature \cite{2020RLWholeHVAC}.}
%    \label{fig:scatter_plot_E_M}
%\end{figure}

\subsection{Total training time}
Besides the training time per episode, we will now investigate on the first aspect for longer training, i.e. on reducing the required number of episodes until convergence of the Q-function (i.e. the DDQNs) is reached.

\paragraph{Layer normalization}
Layer normalization has a strong effect on the effectiveness of the training.
We trained the model with exclusively one room controlled and compared a setting with and without layer normalization, both trained for the same number of episodes.
Without layer normalization the sum of complaint magnitudes over one week is $21.1\%$ higher compared to the setting with layer normalization.
Energy consumption rises by $3.6\%$ compared to the setting with layer normalization.
Other training runs for the same scenario show the same effect.
Thus we can conclude, that layer normalization reduces the number of periods required for training.

\paragraph{Reward scaling}
Scaling the reward to the range $[-1,0]$ seems to have an effect, especially if no layer normalization is used.
The number of required episodes can be reduced by ca. 20\% when using a reward in the range $[-1,0]$.
We manage to get the reward into the range $[-1,0]$ at least most of the time by reducing $\lambda_E$ and $\lambda_M$ with the same factor, so that the ratio of both values stays the same.
Flattening the reward as done in \eqref{eq:e_flat} further reduces the outliers.

%%The impact of reward scaling can be explained by the following example.
%Even when the reward is always within the range $[-1,0]$, there are situations in which the neural network has to predict a wider range of Q-values.
%This is caused by the fact, that the Q-network approximates the Q-function, i.e. the expected, discounted return for a state-action pair (see \eqref{eq:2_bellman_opti_q}).
%If we assume we have a Q-value of $-1$ for all next actions, a discount factor of $\gamma = 0.9$ and a current reward of $r_t = -1$, the new Q-value is $r_t + \gamma \cdot (-1) = -1.9$.

%We can summarize, that we should add a small scaling factor if the rewards are far away from the range $[-1,0]$. Otherwise no work has to be done.

\paragraph{DDQN vs. DQN}
If we train the agents as described above using pretraining, but replace the DDQN-algorithm by just using a DQN, the energy consumption for the example week rises by ca. $16\%$.
The manual setpoint changes show less change.
If we take a closer look to the actions of the agents trained with the DQN algorithm compared to those trained with DDQN, we see that DQN-trained agents tent to produce more volatile actions.

%\paragraph{Epsilon decay}
%We cannot identify changing performance measures in the mean for settings with linear or exponential epsilon decay.
%Using no epsilon decay and thus using a constant value of $0.05$ leads to no real learning.
%Furthermore, if the final epsilon value of $0.05$ is reached but we continue to train the agents with a constant epsilon of $0.05$, then it seems that the agents do not really learn new things.

%\paragraph{Time resolution}
%Experiments with a smaller time resolutions than one hour (e.g. 15 minutes) resulted in a bad convergence of the Q-values.
%Even more, the agents tend to change the heating setpoints very often.
%% our explanation:
%At a time resolution of one hour one agent has 11 actions from which he can choose.
%If we view a time resolution of 15 minutes from the one-hour perspective, every agent has four times as much possible choices so to say.
%Thus it requires much more training time.
%As training already takes long and results are acceptable, we agreed to use a resolution of 1 hour.

\section{Results}
\label{sec:restults}
We now take a detailed look at the concrete energy saving potential and the thermal comfort produced by the framework for heating the example building.

\subsection{Performance measures}
As a reference week which will be used for evaluation we define the first 7 days in January.
We use the components of the reward as defined in \eqref{eq:reward} as the two main performance measures.
These are the sum of consumed energy $E^{total} = \sum_{t} \sum_{d\in D} E_t(d)$ and the sum of manual setpoint complaint magnitudes $M^{total} = \sum_{t} \sum_{z\in Z}\bigtriangleup_t(z)$.
Besides that we take a look at the run time.

\subsection{Energy savings and thermal comfort}
\begin{figure}[htbp]
	\centerline{\includegraphics[scale=1, width=0.65\textwidth]{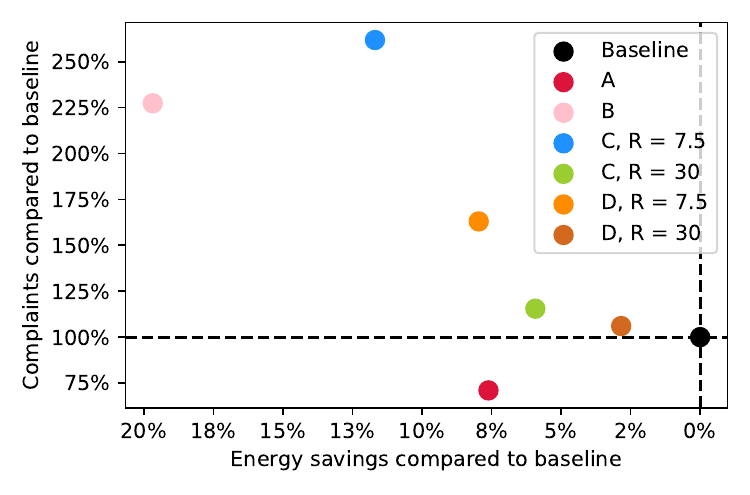}}
    \caption[Scatter plot of energy savings and complaint magnitudes]{Scatter plot showing mean energy savings and complaint magnitudes compared to the baseline for the evaluation week. Different points denote different settings, as refered in the text. The R-value is the ratio $\lambda_m / \lambda_e$. The baseline tiles the plot in four tiles, where the lower left is the region were we save energy and reduce the complaints. The upper left it the region where we save energy, but notify more complaints.}
    \label{fig:scatter_plot_E_M}
\end{figure}

We use the reward with the modification from \eqref{eq:e_flat} and a clip level $l_{clip}$ of $150$ kWh.
All results we present in the following are the average value over more training runs with different random initialization.
In \eqref{eq:reward} we add two factors $\lambda_e$ and $\lambda_m$ for weighting the two reward components, i.e. energy consumption vs. complaints.
If one of these two components overrules the other, we only optimize this one. So finding a balance of the ratio $\lambda_m / \lambda_e$ is essential.
We obtain acceptable results by setting $\lambda_m / \lambda_e$ into the range from 7.5 to 30.
An overview of the following results can be found as scatter plot in Fig. \ref{fig:scatter_plot_E_M} showing the dependency between average energy savings and complaints compared to baseline.

When using no pretraining (Point B in Fig. \ref{fig:scatter_plot_E_M}) and setting the weighting factors of the reward function to a ratio of $\lambda_m / \lambda_e = 7.5$ ($\lambda_e = 0.008$ and $\lambda_m = 0.06$), we can reduce the energy consumption by 19.7\% compared to our baseline.
A drawback is that $M^{total}$ is more than twice the value of the baseline, i.e. thermal comfort is not really met.

When using the single agent in broadcasting mode as pretraining technique (Points C in Fig. \ref{fig:scatter_plot_E_M}) and setting the weighting factors to the same ratio of $\lambda_m / \lambda_e = 7.5$, we can reduce the energy consumption by 11.6\% compared to our baseline.
In this configuration, $M^{total}$ is even 160\% above the baseline value.
When changing the ratio of $\lambda_m / \lambda_e$ to $30.0$ ($\lambda_e = 0.004$ and $\lambda_m = 0.12$) we can reduce $M^{total}$ to be at least 15\% above the baseline value and still get a reduction of energy consumption by around 6\%.

When using a partially rule-based control as pretraining technique (Points D in Fig. \ref{fig:scatter_plot_E_M}) and setting the weighting factors to the first evaluated ratio of $\lambda_m / \lambda_e = 7.5$, we can reduce the energy consumption by 8\% compared to our baseline while increasing the complaint measure $M^{total}$ by 63\%.
When taking a ratio of $\lambda_m / \lambda_e = 30.0$ again, we can reduce $M^{total}$ to be only 6\% above the baseline value and still get a reduction of energy consumption by 2.8\%.

We will now compare our framework with one of the related publications \cite{2020BrandiDRLIndoorTempOpti}.
As their framework optimizes the heat for a single zone only, we make a last evaluation where the heating setpoint for Space 2 is controlled by an RL-agent, whereas the other zones have a rule-based control for the complete time of the training.
%In this setting (Points A in Fig. \ref{fig:scatter_plot_E_M}) we can reduce $M^{total}$ by 30.0\% compared to the baseline and even save ca. 2\% energy on building level.
%As only one thermal zone is controlled by an agent, this is ca. 8\% energy saving for one zone.
%This result is in the same region as the results of \cite{2020BrandiDRLIndoorTempOpti}.
In this setting (Points A in Fig. \ref{fig:scatter_plot_E_M}) we can reduce $M^{total}$ by 30.0\% compared to the baseline and even save ca. 8\% energy for this controlled zone.
So our framework produces results that are comparable to \cite{2020BrandiDRLIndoorTempOpti}.

%\subsection{Summary of the results}
Summarizing these results we can say, that our framework is able to reduce the energy consumption required for heating by setting the heating setpoint for every zone in an intelligent way while fulfilling the wishes of the occupants.
Settings with a ratio of $\lambda_m / \lambda_e = 30.0$ seems to be the best compromise with energy savings around 6\% when using broadcasting pretraining.
Even though the measure of the complaints $M^{total}$ is 14\% or 6\% above the baseline, a closer look at the points of discomfort shows that these slight deviations are acceptable and only occur in the rooms with unsteady occupancy or with occupants that change their comfort temperature during the day.
Using pretraining in combination with shared parameters returns the best results.
%
% TODO: hier könnte man grafik 4.21 anbringen, zu mindestens für zwei Räume
%

\section{Conclusion and Outlook}
\label{sec:conclusion}
In this work we build a RL-framework for optimizing the control of the heating components of a HVAC system.
Every room has one RL-agent with a DDQN as Q-function approximator.
To speed up the training we share the parameters of the networks among the agents and apply pretraining.
The goal of optimization is to reduce energy consumption while maintaining thermal comfort.
The latter is measured by the complaints of modeled occupants.
We define the reward for RL on building level.
Finally, we can increase the sustainability of buildings by reducing the energy consumption by ca. 6\%.
We can also increase the sustainability of developing RL-based control algorithms for HVAC by reducing the required training time thanks to parameter sharing.

At the end of the work, there remain some open questions.
The training should be expanded for a longer period than one month, and thus might require cooling as well.
On top of that, classical HVAC systems have much more components that can be controlled, like the air volume stream or the central boiler. These components are good candidates for an expansion of the framework.
Another aspect will be the evaluation of generalizability of our framework in order to test the transferability to other buildings.
%While generalizability is one of our goals in designing the framework, we have not tested it enough to publish the results yet.

Other RL algorithms like proximal policy optimization (PPO) seems to be promising in the area of controlling HVAC devices.
Using this framework as a basis, a multi-agent variant of PPO will be investigated in future research.
Besides that, further comparison of our framework with MPC is planed.
% we are comparable to MPC, see 2014HVAC_with_MPC with 9.4\% savings

\section*{Acknowledgment}
The authors thanks Niklas Ebell and Oliver Birkholz for the fruitful discussions.

%\section*{References}
\bibliographystyle{ieeetr}
\bibliography{Literatursammlung}

\end{document}